\documentclass[aps,prl,twocolumn,amsfonts,showpacs]{revtex4}
\usepackage{epsfig,amsopn,color}
\usepackage{graphicx}
\usepackage{amsmath}

\begin{document}

\title{Cooling classical particles with a microcanonical Szilard engine}

\author{Rahul Marathe\footnote{Electronic address: rahul@fis.ucm.es}}

\author{J. M. R. Parrondo\footnote{Electronic address: parrondo@fis.ucm.es}}

\affiliation{Departamento de F\'{\i}sica At\'{o}mica, Molecular y
Nuclear and GISC, Universidad Complutense de Madrid, 28040-Madrid,
Spain}

\date{\today }

\begin{abstract}
The possibility of extraction of energy from a system in a cyclic
process is discussed. We present an explicit example where a system,
initially prepared in a microcanonical state, is able to perform
such operation. The example is similar to the Szilard engine, but
the microcanonical initial condition allows one to design a
protocol where measurement is not necessary.
\end{abstract}
\maketitle

Extracting energy from a system faces limitations imposed by
Thermodynamics and Statistical Mechanics. Kelvin statement of the
Second Law precludes the cyclic extraction of energy from a single
thermal bath. Using Hamiltonian dynamics, Jarzynski
\cite{Jarzynski:1997p1857} proved that this is the case for systems
initially prepared in a Boltzmann state. Suppose a system described
by a Hamiltonian $H(q,p;\lambda_1,\dots,\lambda_n)$, where $(q,p)$
stands for a point in the phase space and
$\lambda_1(t),\dots,\lambda_n(t)$ are external parameters operated
by an external agent. The system is isolated, except for the
interaction with this external agent, and consequently the
probability density $\rho(q,p;t)$ obeys the time-dependent Liouville
equation. Jarzynski \cite{Jarzynski:1997p1857} proved that, along an
arbitrary cyclic process $\lambda_i(0)=\lambda_i(\tau)$, the average
energy of the system necessarily increases, i.e., cyclic energy
extraction is not possible. Campisi \cite{campisi2} further 
generalized this result proving that the extraction is impossible
for any initial distribution $\rho(q,p;0)$, such that the resulting
probability density for the energy is a decreasing function. Both
Jarzynski's and Campisi's results can be considered as mechanical
proofs of the Kelvin statement of the Second Law.

On the other hand, the microcanonical distribution, where the total
energy of the system is known with a given accuracy, does not
fulfill Campisi's requirement. In fact, Sato \cite{sato} devised a
particular example ---a one-dimensional particle in a potential
depending on one parameter $\lambda(t)$--- where, starting in a
microcanonical state, the energy of the particle decreases along a
process with $\lambda(0)=\lambda(\tau)$. Although this result is
remarkable, the extraction of energy cannot be repeated, since after
the first cycle the probability density is no longer microcanonical.

In this Letter we present an explicit example where energy can be
extracted in a systematic way. It is a microcanonical version of the
Szilard engine \cite{SLeff:2003p817}, consisting of a classical
particle in a time-dependent potential. In contrast with the
original Szilard model, the microcanonical initial condition allows
one to operate the engine without measurements. Our example provides
a new strategy for microscopic cooling of classical isolated
particles, but also touches some fundamental issues regarding the
mechanical rationale of the Second Law.

\begin{figure}
  \vspace{1cm}
  \includegraphics[width=0.5\textwidth]{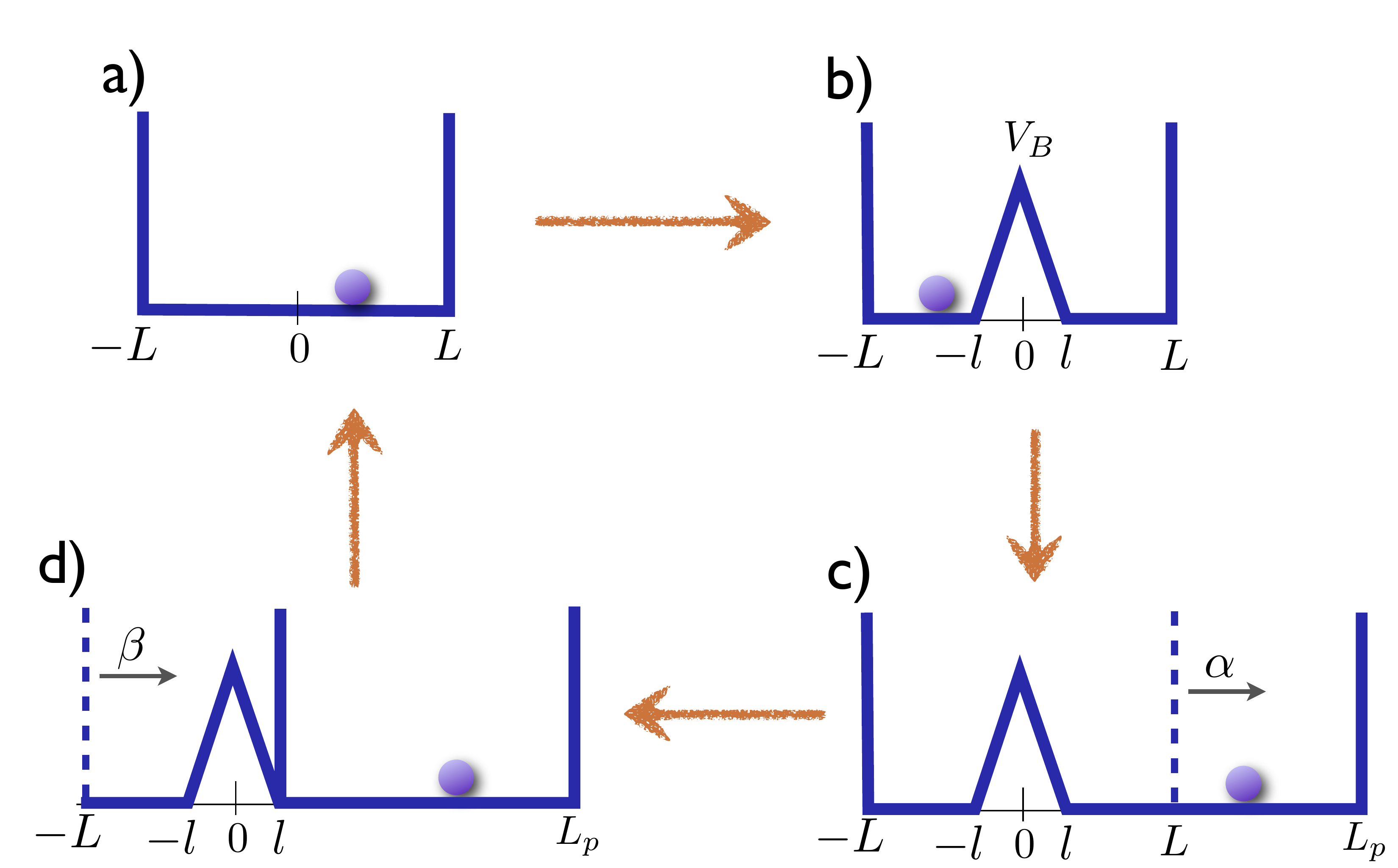}
  \caption{(color online). Different steps in the microcanonical Szilard engine. At stage (b),
 we have plotted the particle in the left half but recall that, under the protocol described in
 the text, the particle may have enough energy to move all over the whole box.} \label{Proto}
\end{figure}

Our  microcanonical version of the Szilard engine
\cite{SLeff:2003p817} consists of a single one-dimensional isolated
particle of mass $m$  in a potential modified by an external agent.
The particle is initially confined in a box of length $2L$ and the
potential is changed as shown in Fig.~\ref{Proto}. In (a--b), a
barrier is raised at the center of the box up to height $V_B$ at
speed $\gamma$. The barrier potential is given by:
\begin{equation}
V(x;t) =
\begin{cases}\displaystyle \frac{\gamma t}{l}\ (l-|x|) &  |x| \le l\\
  0 & \text{otherwise,}
\end{cases}
\end{equation}
for $t\in [0,V_{B}/\gamma]$. 
In the second step (b--c), the right
wall is moved from $L$ to $L_p=2L+l$ with velocity $\alpha$.
Finally, along (c--d), the left wall is moved from $-L$ to $l$, with
velocity $\beta$ and the potential recovers its initial
configuration (but shifted to the right).

\begin{figure}
\includegraphics[width=0.5\textwidth]{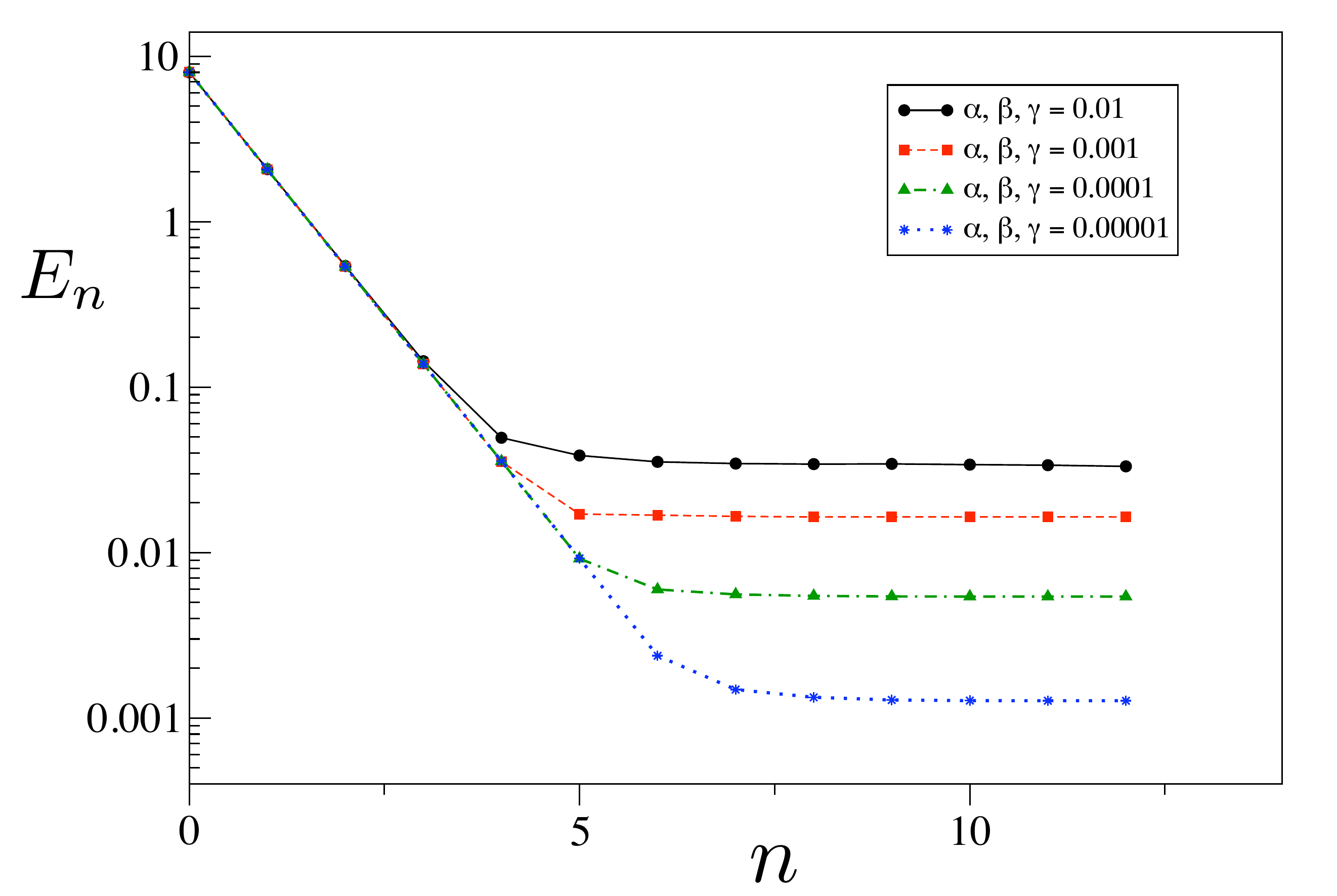}
  \caption{(color online). Log-lin plot of the energy $E_{n}$ 
    versus the cycle number $n$. Parameters used are $m=1$, $L=20$,
    $l=1.0$, for different values of $\alpha$, $\beta$, $\gamma$.
    The maximum height of the barrier in cycle $n$ is $V_B^{(n)}=\kappa^n E_0$,
    with $\kappa=\left[ 1/2+l/(6L)\right]^2$.} \label{xvfEn}
\end{figure}

Now suppose that the barrier height $V_{B}$ is equal to the initial
energy of the particle $E_{0}$. When the barrier is raised along
step (a--b), the particle gains some small amount of energy, and
then the barrier is not high enough to confine the particle in one
of the halves of the container. Along the expansion (b--c), the
particle loses energy in each collision with the right moving wall
(in each collision the velocity changes from $v$ to $-v+2\alpha$).
After a certain number of collisions, its energy drops below $V_{B}$
and the particle gets confined in the rightmost half of the
container. As a consequence, the final compression (c--d) does not
affect the particle much and the net effect of the process is a
reduction of its energy. Since the particle will be {\em always}
confined in the rightmost half of the box, no measurement is
necessary along the step (b--c), unlike the original Szilard engine.
This situation is possible only for certain values of $V_{B}$
depending on the initial energy of the particle, although not on its
position. We can therefore design a protocol that works for
microcanonical initial conditions: fixed energy (velocity) and
random position inside the box. Moreover, if the above steps are
carried out quasi-statically, the energy change is deterministic.
Then, by taking the appropriate values of $V_{B}$ in each cycle, we
can systematically extract energy at least down to a certain value
which can be made arbitrarily small by reducing the speeds $\alpha$,
$\beta$, and $\gamma$. Fig.~\ref{xvfEn} shows the energy after each
cycle, following the described protocol, for different values of
$\alpha,\beta,\gamma$. Data presented in Fig.~\ref{xvfEn} have been
obtained by an event driving algorithm and solving exactly the
Newton's equations for the motion of the particle. Along step (a-b),
a third order polynomial must be solved for the motion along the
barrier.

For a full understanding of our microcanonical Szilard engine, it is
convenient to study in detail the quasi-static limit,
$\alpha,\beta,\gamma\to 0$. Consider a point $(x_{0},p_{0})$ in
phase space, with position $x_{0}$, momentum $p_{0}$, and energy
$E_{0}\equiv H(x_{0},p_{0};0)=p_{0}^2/(2m)+V(x_{0};0)$, evolving as
$(x(t),p(t))$ with energy $E(t)=H(x(t),p(t);t)$. The action
\begin{equation}
\phi_{t}(E(t))\equiv \int_{H(x,p;t)<E(t)} dx\,dp
\end{equation}
is an {\em adiabatic invariant}, i.e., is constant for quasi-static
changes of the Hamiltonian
\cite{Jarzynski:1992p286,TENNYSON:1986p708}. In our case, the
initial action $\phi_{0}=4L\sqrt{2mE_{0}}$ would in principle remain
constant along the cycle. However, the invariance has an exception:
when an orbit changes abruptly due to segregation, confinement or
sudden expansion, induced by the barrier, the action changes
accordingly \cite{TENNYSON:1986p708,CARY:1988p723}. Below we derive
the conditions determining these changes.

For a generic height $V$ of the barrier and location of the left
($L_{l}$) and right ($L_{r}$) walls, the action of an orbit with
energy $E$ reads:
\begin{equation}
\phi(E)=\begin{cases} \displaystyle \sqrt{2mE}\left( 2(L_{l,r}-l) + \frac{4lE}{3V}\right)  & \text{if $E < V$} \\ \\
  \displaystyle \sqrt{2mE}\,\Big\{ 2(L_{r}+L_{l}-2l)\\
 \displaystyle \left. +  \frac{8 lE}{3V}\left[1-\left(1-{V}/{E}\right)^{3/2}\right]\right\} & \text{if $E > V$}
\end{cases}
\label{eq:action}
\end{equation}
where $L_{r}$ or $L_{l}$ in the first line is taken depending on the
side of the box where the orbit is confined. The discontinuity of
the action $\phi(E)$ given by Eq.~(\ref{eq:action}) at $E=V$ is a
trademark of the breaking of its invariance. Suppose that the
barrier is high enough to split, at some stage of step (a--b), the
orbit of a particle with initial energy $E_{0}$. Right before the
segregation, the energy $E_{\rm seg}$ of the particle is equal to
the height $V$ of the barrier. Setting $E=V$ in
Eq.~(\ref{eq:action}) for $E>V$, the invariance for the action
(right before segregation) implies:
\begin{equation}
\sqrt{2mE_{\rm seg}}\left[4(L-l)+\frac{8l}{3}\right]=4L\sqrt{2mE_{0}}
\end{equation}
Segregation actually occurs if the maximum height $V_{B}$ is greater than $E_{\rm seg}$, i.e., if
\begin{equation}
E_{0}<\left(1-\frac{l}{3L}\right)^2V_{B}
\label{lowbound}
\end{equation}
Segregation decreases the action of individual orbits by
\begin{equation}
\Delta \phi_{\rm seg}=-\phi_{0}/2=-2L\sqrt{2mE_{0}}
\label{phiseg}
\end{equation}
Orbits lying in the right side of the box do not suffer any further
collapse or expansion. On the other hand, those in the left hand
undergo a sudden expansion when reaching an energy $V_{B}$ along
step (d--a). The increase of the action equals the action
corresponding to a new lobe added to the orbit in the right side,
which is given by the right contribution in (\ref{eq:action}) at
$E=V=V_{B}$ ($E<V$) and $L_{r}=2L+l$, yielding:
\begin{equation}
\Delta \phi_{\rm exp}=4\sqrt{2mV_{B}}\left( L +\frac{l}{3}\right).
\label{phiexp}
\end{equation}

\begin{figure}[h!]
   \centering
   \includegraphics[width=7cm]{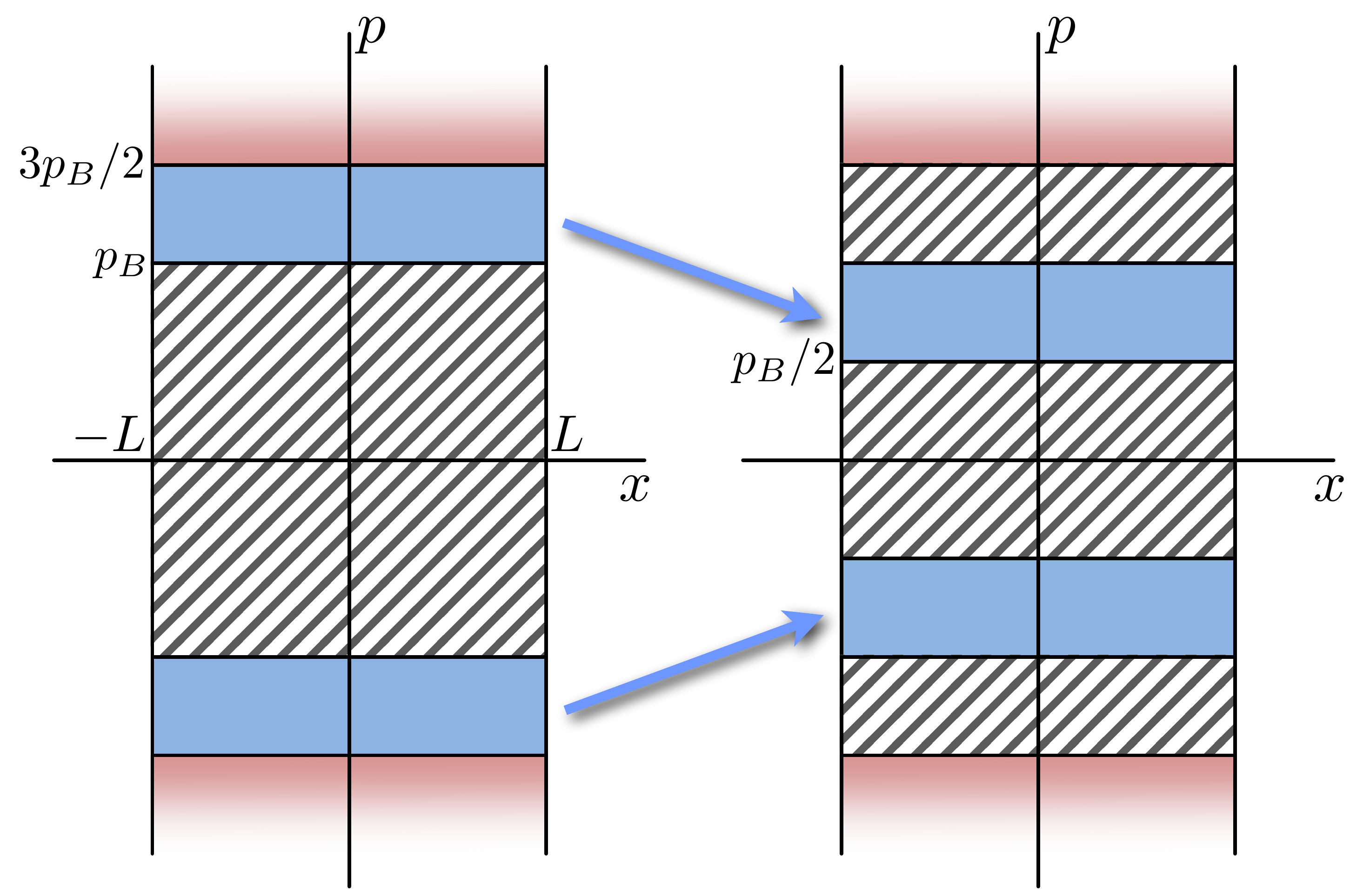}
     \caption{(color online). Evolution of regions in
     phase space after completing the
     protocol with a barrier height $V_B$ and $l=0$, for a single one-dimensional particle.
  The striped region with energy lower than $V_B$, $p\in [-p_B,p_B]$, with $p_B=\sqrt{2mV_B}$,
  splits into three regions leaving room to the blue region $p\in[p_B,3p_B/2]$ to
  move down to $p\in [p_B/2,p_B]$.}
   \label{fig:phasespace}
\end{figure}

If there is no segregation,
i.e., if the particle reaches stage (b) with an energy greater than
$V_{B}$, then along step (b--c) the energy of the particle
decreases, due to collisions with the right moving wall, and
eventually can reach the critical value $V_{B}$ confining the
particle in the rightmost half of the container. In this case, the
orbit of the particle suddenly loses the left lobe (see Fig.
\ref{fig:phasespace}). Before the confinement, the action is
invariant. Setting $E=V=V_{B}$, $L_{l}=L$ in Eq.~(\ref{eq:action})
for $E>V$, the invariance for the action, right before confinement,
reads:
\begin{equation}
\sqrt{2mV_{B}}\left[2(L_{\rm conf}+L-2l)+\frac{8l}{3}\right]=4L\sqrt{2mE_{0}}.
\end{equation}
We have written $L_{r}=L_{\rm conf}$, the position of the right wall
in the moment of confinement, which is at most $L_{p}=2L+l$.
Therefore, confinement occurs if:
\begin{eqnarray}
\sqrt{2mV_{B}}\left[2(3L-l)+\frac{8l}{3}\right] > 4L\sqrt{2mE_{0}}.
\label{upbound}
\end{eqnarray}
Thus Eq.~(\ref{lowbound}) and Eq.~(\ref{upbound}), constitute the condition for
the confinement given by:
\begin{equation}
\left(1-\frac{l}{3L}\right)^2V_{B}<E_{0}<\left(\frac{3}{2}+\frac{l}{6L}\right)^2V_{B}.
\label{condition}
\end{equation}
The change in the action due to confinement is the action corresponding to the left branch of
the orbit disappearing after confinement, which is given by the left contribution in
(\ref{eq:action}) at $E=V=V_{B}$ ($E<V$) and $L_{l}=L$, yielding:
\begin{equation}
\Delta\phi_{\rm conf}=-\sqrt{2mV_{B}}\left(2 L - \frac{2l}{3}\right).
\label{deltaphiconf}
\end{equation}

Finally, if the initial energy is large, $E_{0}>\left[3/2+l/(6L)\right]^2V_{B}$, the orbit spans all
over the box along the whole process and the action is invariant. Summarizing, the final action is
\begin{equation}
\phi_{\rm f}=\phi_{0}+\Delta \phi,
\end{equation}
with $\Delta\phi=\Delta\phi_{\rm conf}<0$
if Eq.~(\ref{condition}) is satisfied, $\Delta\phi=0$ if the energy is above the upper limit
in Eq.~(\ref{condition}), and $\Delta\phi=\Delta\phi_{\rm seg}<0$  or
$\Delta\phi=\Delta\phi_{\rm seg}+\Delta\phi_{\rm exp}>0$ if it is
below the lower limit. These two values are taken with probability $1/2$, respectively.
We can now calculate the final energy $E_{\rm f}$ since
$\phi_{\rm f}=\phi_{0}+\Delta\phi=4L\sqrt{2mE_{\rm f}}$, yielding
\begin{equation}
E_{\rm f}=\left(\sqrt{E_{0}}+\frac{\Delta\phi}{4L\sqrt{2m}}\right)^2.
\label{eq:ef}
\end{equation}
\begin{figure}
  \includegraphics[width=5.5cm]{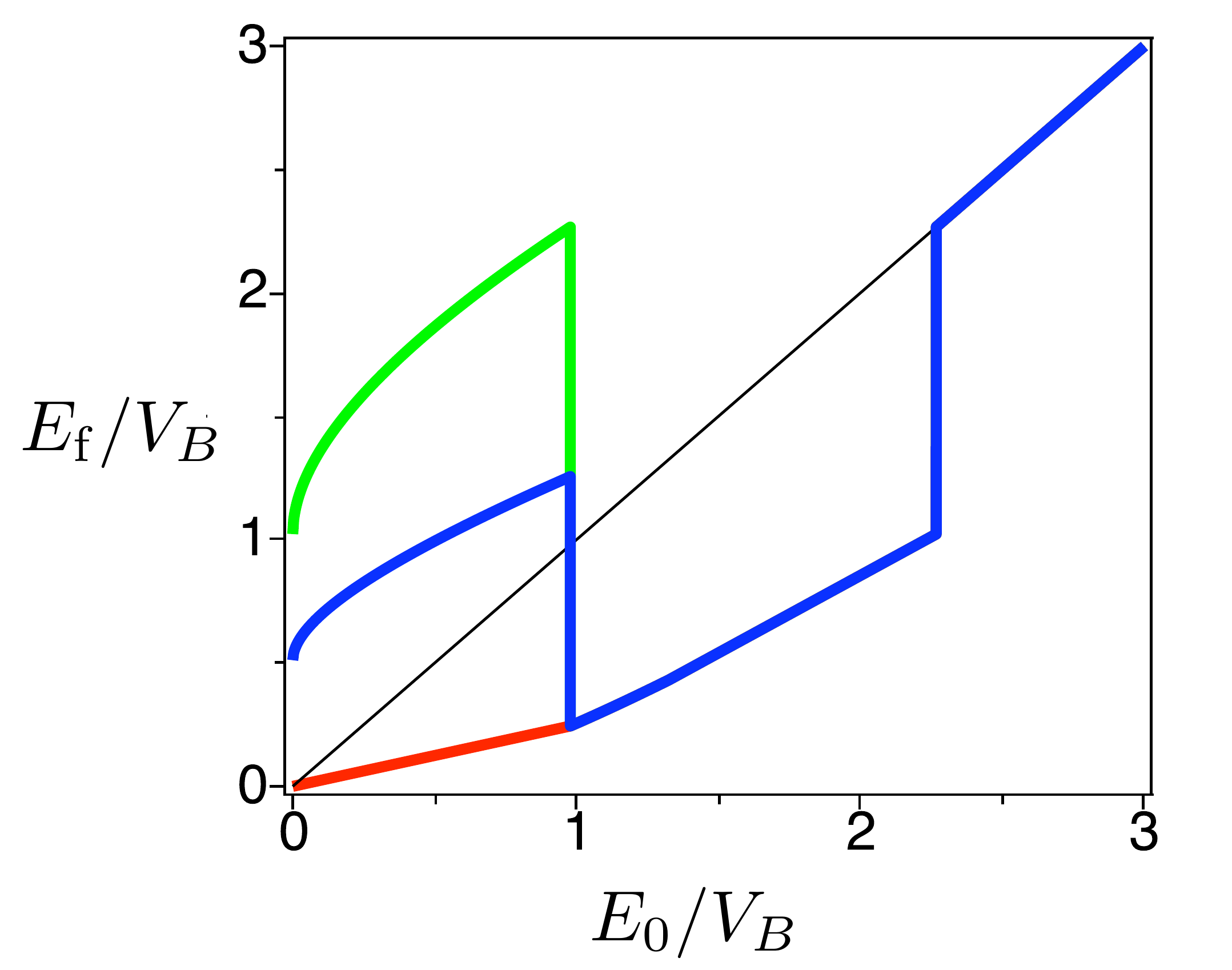}
  \caption{(color online). Final energy as a function of the initial energy in the quasistatic limit,
as given by (\ref{eq:ef}). Parameters used are $m=1$, $L=20$,
$l=1.0$. For small initial energies, particles are segregated
to the left or right half of the box at step (a--b) with probability $1/2$ corresponding,
respectively, to the green and red curves depicted in the figure. The blue curve is
the average final energy.} \label{fig:ef}
\vspace{-0.5cm}
\end{figure}
Fig. \ref{fig:ef} shows the final energy as a function of the initial energy.
We see that the extraction of energy is only possible in the window defined by
Eq.~(\ref{condition}). However, since the energy change in the quasi-static limit is
deterministic  (if there is no segregation) we can tune $V_{B}$ to verify  Eq.~(\ref{condition})
in subsequent cycles. The simplest choice is to set the height of the barrier at cycle
$n$, $V_{B}^{(n)}$, equal to the energy of the particle when the cycle starts.
In the case $V_{B}=E_{0}$, the final energy after a cycle reads $E_{\rm f}=\kappa E_{0}$
with $\kappa=\left[ 1/2+l/(6L)\right]^2$. Therefore, an appropriate choice  is
$V_{B}^{(n)}=\kappa^nE_{0}$, which is the protocol followed  in Fig.~\ref{xvfEn}.
Notice that this choice is made {\em a priori} and does not depend on the actual
evolution of the system. Consequently, for finite velocities $\alpha,\beta,\gamma$
the protocol fails, since the energy does not follow a deterministic sequence as we
depart from the quasi-static limit. However, the energy at which the protocol starts
to fail can be made arbitrarily small by reducing the speed of the process,
as Fig.~\ref{xvfEn} indicates.

Our example prompts a crucial question: is the possibility of
extraction of energy a special feature of one dimensional systems or
can be extended to systems with several degrees of freedom? A naive
extension of the microcanonical Szilard engine to two or more
independent particles does not allow the systematic extraction of
energy. Consider two one-dimensional particles confined in the
interval $[-L,L]$. They are independent except for the initial
preparation in a microcanonical state with total energy $E_0$,
$\rho_{E_0}(p_1,p_2;0)=\delta \left (E_0-
p_1^2/2m-p_2^2/2m\right)/(2\pi m)$. For simplicity, we will assume
that the potential only affects particle 1. Then we can apply the
above results using as initial distribution of  energy the
corresponding distribution of the energy of particle 1:
\begin{eqnarray}
\rho(E;0) &=& \int dp_1\,dp_2\,\delta\left(E-\frac{p_1^2}{2m}\right)\rho_{E_0}(p_1,p_2;0)
\nonumber \\ &=& \frac{1}{\pi\sqrt{E(E_0-E)}}
\end{eqnarray}
\begin{figure}
   \centering
   \includegraphics[width=8cm]{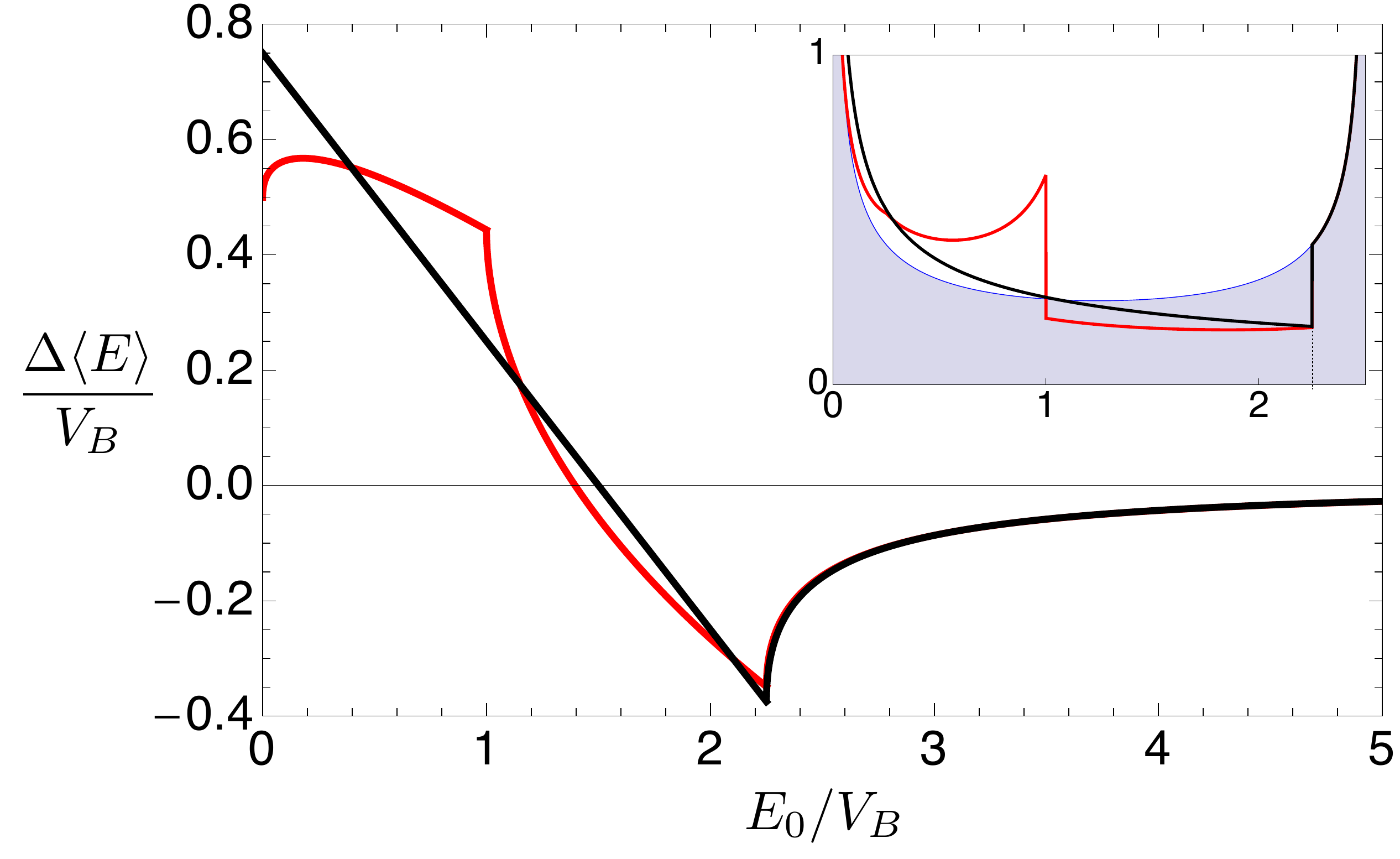}
     \caption{(color online). Average change
     of energy after one (red) and an infinite
     number (black) of repetitions of the protocol with fixed $V_B$ and $l=0$ for two particles,
     as a function of the initial total energy $E_0$. Inset: probability distribution
     of the energy of the particle undergoing the protocol initially (filled blue), 
     after one (red) and an infinite number of repetitions (black)
     for $V_B=1$ and $E_0=2.5$.}
   \label{fig:twoparticles}
\end{figure}

Fig.~\ref{fig:twoparticles} shows the change of energy for an
infinitely narrow barrier $l=0$, as a function of $E_0/V_B$. The red
curve corresponds to the change after one run of the protocol. One
can extract energy in a single run for $E_0\geq 1.4 V_B$, but the
probability distribution of the energy of particle 1 changes in a
rather uncontrolled way (see inset of Fig.~\ref{fig:twoparticles},
red curve), and a repetition of the protocol does not further
decrease the energy. The black curve in Fig.~\ref{fig:twoparticles}
shows the change of energy after an infinite number of cycles (with
the same $V_B$). Adaptive protocols changing $V_B$ in each cycle can
slightly improve the energy decrease, but the systematic extraction
of energy seems impossible in the case of two particles. With three
or more particles, the probability distribution of the energy of
particle 1 is a decreasing function and, following Campisi's theorem
\cite{campisi2}, the energy increases even after the first run of
the protocol.

However,  the basic mechanism of our microcanonical Szilard engine
is a ``shuffling'' of the phase space that could in principle work
in systems with several degrees of freedom.
Fig.~\ref{fig:phasespace} shows how the phase space changes after
one run of the protocol for $l=0$ and a single one-dimensional
particle. Points with energy below $V_B$ (striped regions) are
mapped to two regions of half size; those with energy between $V_B$
and $9V_B/4$ (blue regions) shift down, whereas the rest are not
affected by the protocol. Notice that this shuffling  is fully
compatible with Liouville theorem: the volume of any subset in the
phase space is conserved. On the other hand, it needs the collapse
or split of orbits. Otherwise, the adiabatic invariance of the
volume enclosed by an energy shell implies that every point in the
phase space goes back to its initial position after any quasistatic
cycle.

Up to our knowledge, there is no fundamental obstacle to reproduce
this shuffling of the phase space in a system with many degrees of
freedom. It is an open question to find an explicit example
(probably more involved than a naive generalization of our protocol)
or, on the contrary, to prove that the phenomenon described in this
Letter is exclusive to systems with one degree of freedom. Either
one or the other, the answer to this question touches a fundamental
problem: the mechanical rationale of the Kelvin statement of the
Second Law  for systems prepared in the microcanonical state. Let us
finally recall that the microcanonical ensemble, although of limited
use in real applications, is essential for an objective formulation
of Statistical Mechanics \cite{popescu,reimann1}.

This work has been funded by Grant MOSAICO (Spain). We thank the
hospitality of the Max-Planck-Institut f\"ur Physik komplexer
Systeme (Dresden), where part of this work has been done. We thank
 C. Jarzynski for suggesting Fig.~\ref{fig:phasespace}, K. Sekimoto for pointing
 us Ref.~\cite{sato}, A. Malyshev for the
help with Mathlink program, and R. Brito and E. Roldan for fruitful
discussions.

%\bibliography{biball,nopapers}

%==============================================================================
%BIBLIOGRAPHY
%==============================================================================

\end{document}